# A matter of time: publication dates in Web of Science Core Collection

Weishu Liu

wsliu08@163.com

School of Information Management and Engineering, Zhejiang University of Finance and Economics, Hangzhou 310018, Zhejiang, China

**Abstract:** Web of Science Core Collection, one of the most authoritative bibliographic databases, is widely used in academia to track high-quality research. This database has begun to index online-first articles since December 2017. This new practice has introduced two different publication dates (online and final publication dates) into the database for more and more early access publications. It may confuse many users who want to search or analyze literature by using the publication-year related tools provided by Web of Science Core Collection. By developing custom retrieval strategies and checking manually, this study finds that the "year published" field in search page searches in both online and final publication date fields of indexed records. Each indexed record is allocated to only one "publication year" on the left of the search results page which will inherit first from online publication date field even when the online publication date is later than the final publication date. The "publication year" field in the results analysis page and the timespan "custom year range" field in the search page have the same function as that of the filter "publication year" in search results page. The potential impact of the availability of two different publication dates in calculating bibliometric indicators is also discussed at the end of the article.

**Keywords:** Early access; Publication date; Web of Science Core Collection; Science Citation Index; Bibliometric analysis

## Introduction

Recently, Krauskopf (2019) has found a large gap between the number of papers published by the journal Enfermeria Nefrologica and the ones actually indexed by Scopus. Inspired by this study, we search the publications published by the journal Scientometrics and indexed by another widely used database Web of Science Core Collection (Li et al. 2018; Zhu & Liu 2020). We also find that 396, 398, and 332 records are indexed for the "publication years" 2017, 2018, and 2019 respectively (refined by "publication years" in search results page). By referring to the publisher's webpage and counting the number of records published in each issue, we find that 400, 401, and 307 records were published by Scientometrics in 2017, 2018, and 2019 respectively. The numbers of records published by Scientometrics in 2017 and 2018 are similar for two different data sources. However, it is interesting to find that 25 more records were published by Scientometrics in 2019 according to Web of Science Core Collection than the number counted from the journal publisher's website. We are curious about the true meaning of the filter "publication years" in search results page.

Previous studies have investigated multi-versions of publication dates such as online first, issue, and indexation dates (Alves-Silva et al. 2016; Das & Das 2006; Haustein et al. 2015; Maflahi & Thelwall 2018). Besides, the publication delay and early view effects have been widely explored (Al & Soydal 2017; González-Betancor & Dorta-González 2019; Heneberg 2013; Hu et al. 2018; Kousha et al. 2018; Yu et al. 2005). Web of Science Core Collection has also begun to index early access articles with online publication dates from publishers since December 2017 (similar expressions such as in press/online first for different journals)[1]. When the final version is available, these records will get the final publication dates and retain the online publication dates in the database. This study will investigate how Web of Science Core Collection handles the two different publication dates issue when searching and analyzing the literature online.

## Data and methods

This study chooses Web of Science Core Collection and Journal Citation Reports subscribed by Xi'an Jiao Tong University as the data sources. Based on the subscription of citation indexes, Science Citation Index Expanded (SCI-E, 1900-present), Social Sciences Citation Index (SSCI, 1900-present), Arts & Humanities Citation Index (A&HCI, 1975-present), Conference Proceedings Citation Index- Science (CPCI-S, 1996-present), Conference Proceedings Citation Index- Social Science & Humanities (CPCI-SSH, 1996-present), and Emerging Sources Citation Index (ESCI, 2015-present) under the Web of Science Core Collection are selected[2]. The data were accessed on August 2 2020.

## Analyses

### "Year published" field in search page

When searching for literature in Web of Science Core Collection, users often set the publication year to a specific period to locate the target literature more accurately. For example, users can set the "year published" (PY) field to 2018 in basic search to retrieve records published in 2018 (set the timespan defaulted as "all years" throughout this study unless otherwise indicated). Similarly, users can also set the "year published" field as 2018 in advanced search. By setting "year published" field from 2016 to 2020 respectively in both basic search and advanced search, five pairs of identical results can be obtained. That is to say, the function of "year published" field is same in basic and advanced searches.

Interestingly, by setting "PY=2017 and PY=2018" in advanced search, 699 records can be hit. Similarly, by setting "PY=2018 and PY=2019" in advanced search, 1300 records can be hit (Table 1). By checking the search results manually, we find that the "year published" field tag searches in both online publication date and final publication date fields of indexed records. Same conclusion can be got by using the "year published" field in basic search. According to Clarivate, Web of Science Core Collection began to collect early access contents from December 2017[3]. Once one early access record is fully published and indexed, the early access label will be removed. However,

---

[1] https://support.clarivate.com/ScientificandAcademicResearch/s/article/Web-of-Science-Core-Collection-Early-Access-articles?language=en_US
[2] As advised by Liu (2019), we specify the details of citation indexes used in this study for reproducitivity.
[3] https://clarivate.com/webofsciencegroup/wp-content/uploads/sites/2/dlm_uploads/2019/08/WoS527-external-release-notes-final1.pdf

Web of Science Core Collection will retain the online publication date and also add the final publication date. According to Table 1, it seems that the numbers of records with different online and final publication year values are limited especially from the relative perspective. However, the number of hit records for the search "PY=2019 and PY=2020" has reached 80226 when searching on August 2 2020, which is no longer neglectable.

Table 1 "Year published" search field in basic and advanced searches

| Basic search (Timespan=All years) | | Advanced search (Timespan=All years) | |
| --- | --- | --- | --- |
| Search query | Number of records | Search query | Number of records |
| Year Published=2016 and Year Published=2017 | 0 | PY=2016 and PY=2017 | 0 |
| Year Published=2017 and Year Published=2018 | 699 | PY=2017 and PY=2018 | 699 |
| Year Published=2018 and Year Published=2019 | 1300 | PY=2018 and PY=2019 | 1300 |

**"Publication years" in search results page**

After searching in Web of Science Core Collection, the search results page also provides several filters such as "Publication years", "Web of Science Categories", and "Document types", to name just a few, to help users to refine the results. It is interesting to test if the filter "Publication years" ("Publication year(s)" with double quotation marks in this study denotes the filter field in the search results page) here is same to the "Year published" field in search page.

This study first uses the field "Year published" in search page to retrieve records published in 2016, 2017, 2018, and 2019 respectively, and subdivides the results by the filter "publication years" in search results page. The details are demonstrated in Table 2. When searching "PY=2016", all the retrieved records are allocated to the "publication year" 2016 in search results page. Similar phenomenon can be obtained when searching "PY=2017".

Table 2 The "publication years" filter in search results page

| Search query (Timespan=All years) | Number of total records | Subdivided by "publication years" in WoSCC search results page | | | | |
| --- | --- | --- | --- | --- | --- | --- |
| | | 2016 | 2017 | 2018 | 2019 | 2020 |
| PY=2016 | 2936841 | 2936841 | 0 | 0 | 0 | 0 |
| PY=2017 | 3002103 | 0 | 3002103 | 0 | 0 | 0 |
| PY=2018 | 3053457 | 0 | 699 | 3052546 | 212 | 0 |
| PY=2019 | 3143563 | 0 | 11 | 1088 | 3141856 | 608 |

Note: WoSCC, Web of Science Core Collection.

However, when searching "PY=2018" and subdividing the records by the filter "publication years" in search results page, we find that almost all the retrieved records are allocated to the "publication year" 2018. Moreover, 699 records are allocated to the "publication year" 2017 and 212 records are allocated to the "publication year" 2019. The sum of records belonging to these three "publication years" equals the number of records retrieved by "PY=2018". Similar conclusion can be got for the search query "PY=2019". That is to say, even for some records with different values regarding the online and final publication years, each record will be allocated to only one "publication year" in search results page.

How does Web of Science select the "publication year" values for records with different online and final publication year values in the database is worthy of further investigation. For the results of search query "PY=2018", we further refine the results by selecting the "publication year" 2017. 699 records can be hit and all these 699 records' values of final publication years are 2018 but these records' values of online publication years are 2017. That is to say, Web of Science Core Collection will allocate records into "publication years" according to their online publication dates first if online publication dates are earlier than final publication dates. Contrarily, if we refine the search results of "PY=2018" by setting the "publication year" to 2019, we find that all these records are finally published in 2018 but with online publication years of 2019. That is to say, one record will be allocated to one "publication year" in search results page first according to this record's online publication date field if this value is available (even when the final publication date is earlier than the online publication date). The conclusion holds for the search queries "PY=2019" and "PY=2020".

The results analysis page also provides the function to analyze the results by "publication years". By examining the data, this study finds that the "publication years" field here has the same function as that in the search results page.

**Timespan: "custom year range" in search page**

Some users of Web of Science Core Collection may also use "custom year range" of timespan to search the literature more accurately[4]. Similarly, we are curious about the meaning of "custom year range" here. In order to probe this question, we design the search queries in Table 3 to uncover the true meaning of "custom year range" of timespan here.

For better investigation, this study introduces the DOI (DO) field tag which will retrieve in the DOI field of indexed records[5]. The DOI should begin with the prefix of "10."[6], therefore this study uses "DO=10.*" to retrieve records with legal DOI values in Web of Science Core Collection.

For the year 2016, identical results can be obtained by setting "PY=2016" or "Timespan=2016" (scenario 2016 in Table 3). It is also true for the year 2017 (scenario 2017). However, different phenomena happened in scenarios 2018 and 2019. For example, when using "DO=10.*" and setting "Timespan=2018" to search, 2399463 records can be hit (search query #9 in Table 3). However, when using "DO=10.* and PY=2018" and setting "Timespan=All years" to search, 2400374 records can be hit which is a bit more than the number of hit records by search query #9 (search query #10 in Table 3). We further refine the results by limiting the "publication year" to 2018 in search results page (search query #11 in Table 3), 2399463 records can be got. By using the search query #12 in Table 3, we find that the results of search queries #9 and #11 are identical. That is to say, the timespan "custom year range" in search page has the same function as the "publication years" in search results page. This finding can be reconfirmed in scenario 2019.

---

[4] More information about the timespan, please refer to
https://images.webofknowledge.com/images/help/WOS/hp_timespan.html
[5] For more information about the DOI search in the Web of Science Core Collection, please refer to Zhu et al. (2019a, 2019b).
[6] For more information about the naming rules of DOI, please refer to
https://www.doi.org/doi_handbook/2_Numbering.html

Table 3 Timespan: "custom year range" in search page

| Scenario | Set | Search query | Number of records |
|---|---|---|---|
| 2016 | #1 | DO=10.*; Timespan=2016 | 2205248 |
| | #2 | DO=10.* and PY=2016; Timespan=All years | 2205248 |
| | #3 | DO=10.* and PY=2016; Timespan=All years; Refined by: Publication years:(2016) | 2205248 |
| | #4 | #1 and #3; Timespan=All years | 2205248 |
| 2017 | #5 | DO=10.*; Timespan=2017 | 2279677 |
| | #6 | DO=10.* and PY=2017; Timespan=All years | 2279677 |
| | #7 | DO=10.* and PY=2017; Timespan=All years; Refined by: Publication years:(2017) | 2279677 |
| | #8 | #5 and #7; Timespan=All years | 2279677 |
| 2018 | #9 | DO=10.*; Timespan=2018 | 2399463 |
| | #10 | DO=10.* and PY=2018; Timespan=All years | 2400374 |
| | #11 | DO=10.* and PY=2018; Timespan=All years; Refined by: Publication years:(2018) | 2399463 |
| | #12 | #9 and #11; Timespan=All years | 2399463 |
| 2019 | #13 | DO=10.*; Timespan=2019 | 2606740 |
| | #14 | DO=10.* and PY=2019; Timespan=All years | 2608447 |
| | #15 | DO=10.* and PY=2019; Timespan=All years; Refined by: Publication years:(2019) | 2606740 |
| | #16 | #13 and #15; Timespan=All years | 2606740 |

**Discussion**

Web of Science Core Collection began to index online first articles with online publication dates from publishers from December 2017. These records will be temporally assigned an additional document type "Early Access" before final publication[7]. After the final version is available, these records will contain both online publication dates and final publication dates. This new update will confuse users who use date-related tools to retrieve and analyze records in Web of Science Core Collection, especially in the case that more and more online first articles are indexed by Web of Science.

In this study, we probe how "year published" and "timespan: custom year range" in search page, and the filter "publication years" in search results page deal with the two publication dates problem. Based on the above analyses, we find that the "year published" field in the search page searches in both online and final publication dates fields of indexed records if two values are available. In the search results and results analysis pages, each record will be allocated to only one "publication year" even when different online and final publication years are available. However, one record will be allocated to only one "publication year" inheriting first from online publication date field even when the online publication date is later than the final publication date. The timespan "custom year range" field has the same function as the "publication years" field in search results page[8].

---

[7] https://support.clarivate.com/ScientificandAcademicResearch/s/article/Web-of-Science-Core-Collection-Early-Access-articles?language=en_US

More and more articles are first indexed by Web of Science Core Collection as "Early Access" and Web of Science updates the metadates when the fully published versions are available. These records will have both online publication dates and final publication dates in the database. Two different dates will affect the calculation of time-related bibliometric indicators if the publication dates are used inconsistently. For example, how to calculate the journal impact factor (JIF) for journals which have some records with different online and final publication year values. Although early access articles are excluded from the calculation of a journal's impact factor[9], we find that Web of Science began to count articles which were published online in 2018 but finally published in 2019 as denominator in the calculation of the latest 2019 JIF[10]. This policy may influence the comparability of JIFs of the same journal in different years and also JIFs of journals in the same Web of Science category but with different early access article index policies (Liu et al. 2018a).

Although Web of Science Core Collection has its own policy to deal with the different publication dates problem, Scientometrics researchers should at least be familiar with its policy and investigate the strengths and weaknesses of different coping strategies. This study uncovers the black box that how Web of Science deals with different publication dates in retrieval and online analysis phases. However, the question that which publication date is more suitable to be used in the calculation of various bibliometric indicators still needs further investigation (Donner 2018; Gai et al. 2015; Tort et al. 2012; Wang 2013).

With the continuous update of Web of Science Core Collection, the database provider should provide more technical details about its update for users timely. Besides, users should also be aware of the features and defects of this authoritative bibliographic database which have been widely investigated in previous studies (Birkle et al. 2020; Franceschini et al. 2016; Huang et al. 2017; Liu et al. 2018b, 2020; Martín-Martín et al. 2018; Mongeon & Paul-Hus 2016; Tang et al. 2017).

**Acknowledgements** This research is financially supported by the National Natural Science

---

[8] A latest study also finds some abnormal phenomena for the year 2000 (Hu et al. 2020). Web of Science Core Collection may update its policies continuously or be with a very small percentage of errors.

[9] https://support.clarivate.com/ScientificandAcademicResearch/s/article/Journal-Citation-Reports-Online-Early-Ahead-of-Print-and-In-Press-Articles?language=en_US

[10] For example, 78 citable items which were published online in 2018 but finally published in 2019 were counted into the calculation of the 2019 journal impact factor of the journal Antioxidants & Redox Signaling. During the proofreading process, we received the feedback from Clarivate as following: "The current policy is for the JCR to use the final publication year. We are aware of a relatively small number of journals, including Antioxidants and Redox Signaling, where articles that were indexed as Early Access in 2018 and final in 2019 were inadvertently included in the denominator. These are being addressed in the JCR reload—the Early Access 2018 items and the citations to those items will be removed from the Journal Impact Factor (JIF) calculation. As a result of this change, we forecast the revised JIF of Antioxidants and Redox Signaling will be 7.040."